\documentclass[12pt,preprint]{aastex}
\begin{document}
\newcommand{\etal}{{\it et al.}}

\title{Discovery of a 115 Day Orbital Period in the Ultraluminous
X-ray Source NGC 5408 X-1} \author{Tod E. Strohmayer}
\affil{Astrophysics Science Division, NASA's Goddard Space Flight
Center, Greenbelt, MD 20771; tod.strohmayer@nasa.gov}

\begin{abstract}

We report the detection of a 115 day periodicity in {\it SWIFT}/XRT
monitoring data from the ultraluminous X-ray source (ULX) NGC 5408
X-1.  Our ongoing campaign samples its X-ray flux approximately twice
weekly and has now achieved a temporal baseline of $\approx 485$ days.
Periodogram analysis reveals a significant periodicity with a period
of $115.5 \pm 4$ days. The modulation is detected with a significance
of $3.2 \times 10^{-4}$. The fractional modulation amplitude decreases
with increasing energy, ranging from $0.13 \pm 0.02$ above 1 keV to
$0.24 \pm 0.02$ below 1 keV.  The shape of the profile evolves as
well, becoming less sharply peaked at higher energies. The periodogram
analysis is consistent with a periodic process, however, continued
monitoring is required to confirm the coherent nature of the
modulation.  Spectral analysis indicates that NGC 5408 X-1 can reach
0.3 - 10 keV luminosities of $\approx 2 \times 10^{40}$ ergs
s$^{-1}$. We suggest that, like the 62 day period of the ULX in M82
(X41.4+60), the periodicity detected in NGC 5408 X-1 represents the
orbital period of the black hole binary containing the ULX.  If this
is true then the secondary can only be a giant or supergiant star.

\end{abstract}


\section{Introduction}

The existence of black holes in the mass range from $10^{2} - 10^{4}$
$M_{\odot}$--intermediate mass black holes (IMBH)--is still not widely
accepted.  It has been argued based on their extreme luminosities that
some of the bright X-ray sources found in nearby galaxies, the
ultraluminous X-ray sources (ULXs), may be IMBHs (Colbert \& Mushotzky
1999), œbut it has also been suggested that these objects may appear
luminous due to beaming of their X-ray radiation (King et al. 2001).
As yet there has been no direct measurement of the mass of a ULX.

At present the best IMBH candidates include X41.4+60, the brightest
ULX in the starburst galaxy M82 (also referred to as M82 X-1, Kaaret,
Feng \& Gorski 2009; Strohmayer \& Mushotzky 2003), and the ULX NGC
5408 X-1 (Strohmayer et al. 2007; Kaaret \& Corbel 2009).  Very
recently, Farrell et al. (2009) have reported the identification of an
X-ray source (2XMM J011028.1-460421) very near the absorption line
galaxay ESO 243-49. They argue for a physical association with this
galaxy which at a redshift of $z = 0.0224$ implies a luminosity in
excess of $10^{42}$ erg s$^{-1}$, and suggest it is an IMBH in excess
of 500 $M_{\odot}$. If the association with ESO 243-49 is correct then
this object is the most luminous ULX currently known.

X41.4+60 is also amongst the most luminous ULXs, on occasion having a
luminosity upwards of $10^{41}$ ergs s$^{-1}$ (Kaaret et
al. 2009). This object also shows a 62 day periodic modulation in its
X-ray flux, which has been proposed to be the orbital period of the
binary system containing the black hole (Kaaret, Simet \& Lang 2006;
Kaaret \& Feng 2007).  Both NGC 5408 X-1 and X41.4+60 show
quasiperiodic oscillations (QPOs) in their X-ray fluxes, and these ULX
QPOs appear at systematically lower frequencies than the analogous
QPOs observed in Galactic black hole binary systems (Strohmayer \&
Mushotzky 2003; Strohmayer et al. 2007).  Indeed, Strohmayer \&
Mushotzky (2009) have recently shown that the QPO properties in NGC
5408 X-1 correlate with the energy spectrum and X-ray flux in a manner
fully consistent with the behavior observed for so-called Type C QPOs
in Galactic systems.  The Type C QPOs in stellar mass black holes are
strong (fractional rms amplitude $\sim 15\%$), relatively coherent
($\nu_{qpo} / \Delta\nu > 10$) oscillations with characteristic
frequencies of $\sim 1 - 10$ Hz that vary in frequency in correlation
with source flux and spectral index (see Sobczak et al 2000; Vignarca
et al. 2003; Casella, Belloni \& Stella 2005). They are associated
with an energy spectral state in which approximately half or more of
the flux is carried by a power-law component with slope $\Gamma \sim 2
- 2.5$, commonly referred to as the Steep Power Law state (SPL,
McClintock \& Remillard 2006), or the Hard Intermediate State (HIMS,
Belloni 2006). The precise origin of these QPOs is still uncertain,
but their phenomenology has been used to estimate the masses of black
holes (see Shaposhnikov \& Titarchuk 2009).

The Type C QPOs in NGC 5408 X-1 identified by Strohmayer \& Mushotzky
(2009) have frequencies of a few tens of mHz, a factor of 100 lower
than the Galactic black holes.  Scaling the observed QPO frequencies
in NGC 5408 X-1 to those observed in Galactic systems of known mass
suggests that NGC 5408 X-1 is an IMBH with a mass greater than 1000
$M_{\odot}$ (Strohmayer \& Mushotzky 2009).  Moreover, recent optical
spectroscopy reported by Kaaret \& Corbel (2009) indicates that the
optical counterpart to NGC 5408 X-1 is associated with reprocessed
emission from an accretion disk as well as a strongly photoionized
optical nebula. This, combined with the presence of a powerful radio
nebula also appears consistent with the presence of an IMBH (Lang et
al. 2007; Soria et al. 2006; Kaaret et al. 2003).

While it is likely that many of the ULXs are accreting black hole
binary systems, very little is known about the nature of these
putative binaries.  If some of the brighter systems are indeed IMBHs
accreting via Roche lobe overflow, then their orbital periods could be
as long as a few hundred days (Portegies Zwart et al. 2004).  With the
exception of X41.4+60 noted above, very few objects have been
monitored with a cadence and duration that would enable sensitive
searches for X-ray periods in the range of 10s to 100s of days.  The
{\it Swift} observatory uniquely provides both the X-ray imaging
sensitivity and scheduling flexibility to enable such searches.
Starting with Observing Cycle 4 {\it Swift} has been monitoring NGC
5408 X-1 several times per week as part of an approved program.  These
observations have continued into Cycle 5 and are presently ongoing.
Here we present results from this campaign which provide evidence for
a 115 day periodicity in NGC 5408 X-1 which could very likely be the
orbital period of a Roche lobe overflow binary containing an IMBH.

\section{Data Extraction and Analysis}

Monitoring of NGC 5408 X-1 with the {\it Swift} XRT began on 2008
April 9. Since then, the source has been observed 2 - 3 times per week
with a typical exposure of 2 ksec.  Aside from an $\approx 80$ day gap
beginning on 2008 September 26, the observing cadence has been
unbroken.  Results presented here include observations through 2009
August 5.

We began our analysis with the Level 2 cleaned XRT event files.  The
XRT was operated in photon counting mode.  We extracted good events
using an extraction radius of 25'' around the position of NGC 5408
X-1.  Most observation sequences were made up of one to two good time
intervals (GTI). In general, we combined these to create a single
count rate estimate for each observation sequence.  An important issue
is to account for bad pixels and columns present on the XRT CCD. The
University of Leicester's XRT Digest webpage (see,
http://www.swift.ac.uk/xrtdigest.shtml\#analysis) provides a detailed
discussion of the problem.  In order to account for the potential loss
of effective exposure due to bad pixels we constructed exposure maps
for each observation using the tool {\it xrtexpomap}.  We then
integrated over the extraction region in the exposure map, weighted by
the XRT point spread function (Moretti et al. 2006).  This provides an
effective exposure for each observation.  We used the corrected
exposures in order to estimate source counting rates for each
observation.  This procedure resulted in a total of 113 count rate
estimates.  The resulting light curve is shown in Figure 1.  The mean
rate is 0.061 s$^{-1}$, and excursions from 0.01 to 0.1 s$^{-1}$ are
evident.

We extracted spectra from each observation using the same extraction
regions as for the count rate estimates.  Background estimates were
extracted from regions of the same size nearby on the CCD.  Previous
observations of NGC 5408 X-1 suggest that the spectrum does not change
dramatically (Kaaret et al. 2003; Strohmayer et al. 2007; Strohmayer
\& Mushotzky 2009), so we combined all observations into a single,
average spectrum in order to obtain an average count rate to flux
conversion.  We fit the resulting spectrum with the sum of a power-law
and disk blackbody (model {\it diskpn} in XSPEC), and fixed the
absorbing column at the value measured previously (Strohmayer et
al. 2007). We find a disk temperature $kT_{max} = 0.18 \pm 0.02$ keV
and power-law index $\Gamma = 2.6 \pm 0.2$.  These values are
reasonably consistent with previous XMM-Newton measurements.  We find
that with this spectral form an XRT count rate of 0.06 s$^{-1}$
corresponds to a 0.3 - 10 keV unabsorbed flux of $4.0 \times 10^{-12}$
ergs cm$^{-2}$ s$^{-1}$.  At a distance of 4.8 Mpc this corresponds to
a luminosity of $1.1 \times 10^{40}$ ergs s$^{-1}$.  The implied peak
luminosity during these observations was $\approx 1.9 \times 10^{40}$
ergs s$^{-1}$.  We quote fluxes and luminosities in the 0.3 - 10 keV
range for comparison with previous measurements, however, we note that
extending the band pass down to 0.1 keV increases these luminosity
estimates by almost a factor of two.

We carried out a periodogram analysis using the methods of Scargle
(1982) and Horne \& Baliunas (1986).  We calculated the periodogram at
140 frequency points, which is the number of independent frequencies
estimated by the method of Horne \& Baliunas (1986), and we normalized
by the total variance in the data. The resulting power spectrum is
shown in Figure 2.  The highest peak in the spectrum appears at a
frequency of $1.0018 \times 10^{-7}$ Hz (corresponding to a period of
115.5 days), and has a value of 24.7.  We estimate the significance of
the peak by calculating the chance probability for obtaining this
value from the $\chi^2$ distribution with 2 degrees of freedom, and
multiplying by the number of trials (140 frequency bins).  This yields
a significance of $6.1 \times 10^{-4}$, which is better than a
$3\sigma$ detection.  There is no evidence for significant power in
any other frequency range. The coherence $Q$, defined as the center
frequency divided by the width of the peak (measured as the FWHM), is
4.7 and is consistent with a periodic modulation.  However, the
present data cover only a bit more than 4 cycles, so further
monitoring will be required to better assess the coherence of the
modulation.

We note that Kaaret \& Feng (2009) have recently submitted a paper to
astro-ph (0907.5415) which reports results using some of the same data
presented here. We became aware of this work after beginning work on
this manuscript.  Kaaret \& Feng do not claim to detect a significant
periodicity in the NGC 5408 X-1 light curve, but they do state that
the highest peak they see is ``near a period of 115 days.''  We
believe the reason for this seeming discrepancy is two-fold. First, we
have used more data in our analysis, that is, a longer temporal
baseline, and secondly, we have included the exposure map corrections
in our light curve.  Indeed, when we restrict our analysis to the time
range used by Kaaret \& Feng, and ignore the exposure corrections,
then we see a drop in the peak power at 115 days to a level consistent
with Kaaret \& Feng. The fact that the exposure correction increases
the Fourier power is further evidence that the 115 day modulation is a
real signal.

To further explore the nature of the modulation we folded the data in
several energy bands at the measured period of 115.5 days.  To do this
we placed each light curve measurement in its appropriate phase bin,
and then we averaged all measurements in a given bin. Errors were
determined by averaging the individual errors in quadrature. The
resulting profiles in twelve phase bins are shown in Figure 3. We show
profiles in the full band ($E > 0.2$ keV, upper left), two soft bands
($E < 1$ keV, upper right; $E < 0.7$ keV, lower left), and a hard band
($E > 1$ keV, lower right), and we plot two cycles for clarity.  We
fit a model to each profile that includes two Fourier components (the
fundamental and first harmonic), $I = A + B\sin 2\pi(\phi - \phi_0) +
C\sin 4\pi(\phi-\phi_1)$.  Results of these fits are summarized in
Table 1. For the hard band ($E > 1$ keV) profile the harmonic term is
not statistically required. The folded profiles reveal interesting
energy dependent behavior. For example, the modulation amplitude
appears to clearly decrease with increasing energy. Defining the
fractional amplitude $f_{amp} = (B^2 + C^2)^{1/2} / A$, we find a
significant change from $0.133\pm 0.018$ to $0.241 \pm 0.017$ in going
from $E > 1$ to $E < 1$ keV.  Indeed, recomputing the periodogram with
only $E < 1$ keV photons results in an increase in the peak Fourier
power to 27.33. The detection significance then improves to $3.2
\times 10^{-4}$, which includes an additional factor of 2 for the
increased number of trials.  In addition to the amplitude variations
the profile appears smoother at higher energies, and the phase of the
fundamental component is significantly different above 1 keV than
below, in the sense that the harder photons appear to lag the soft
photons with a relative phase difference of $0.11 \pm 0.025 = 12.6 \pm
2.9$ days. Finally, we computed the hardness ratio as a function of
phase, where we define the hardness ratio as the count rate for $E >
1$ keV divided by the rate for $E < 1$ keV.  Figure 4 shows the
variation of the hardness with phase, and we have also plotted the
full-band folded profile for comparison (dashed curve). The hardness
ratio shows a sharp rise to a maximum near phase 0.45, followed by a
more gradual decline. The peak of the modulation is clearly softer
than the minimum.

\section{Discussion and Implications}

A number of accreting binaries show X-ray modulations at their orbital
periods. Among black hole binaries Cyg X-1 and Cyg X-3 are well known
examples (Wen et al. 1999; Elsner 1980), and more recent detections
include LMC X-3 (Boyd et al. 2001), 1E 1740.7-2942 and GRS 1758-258
(Smith et al. 2002). Many neutron star binaries also show X-ray
modulations at the binary orbital period.  The orbital modulations in
accreting binaries have typically been attributed to periodic
obscuration produced by a vertically and azimuthally structured
accretion disk or scattering in an extending accretion disk corona or
stellar wind from the donor (Parmar \& White 1988; Wen et al. 1999). A
recent example of an X-ray modulation at the putative orbital period
in a neutron star system is that of GX 13+1 (Corbet 2003). Thus, if
ULXs are indeed accreting black hole binaries, then it is not
unexpected to detect X-ray modulations at their orbital periods.

While their have been several claims of detection of orbital
modulations in ULXs (see, Kaaret, Simet \& Lang 2006 for a brief
summary), the most compelling detection to date is that of the 62 day
modulation from the M82 ULX X41.4+60 (Kaaret, Simet \& Lang 2006;
Kaaret \& Feng 2007).  The average modulation in X41.4+60 as observed
with the RXTE/PCA is roughly sinusoidal with an amplitude of $\approx
20\%$ (Kaaret \& Feng 2007). This is qualitatively consistent with the
$0.2 < E < 8$ keV modulation amplitude and profile that we report here
for NGC 5408 X-1, and suggests that similar physical processes may
produce the observed modulation in both systems.

Superorbital periods are known in both neutron star and black hole
binaries. Well known examples include the 35 day modulation in the
neutron star system Her X-1, and the 164 day period in the putative
black hole binary SS 433 (Wijers \& Pringle 1999).  These variations
have generally been ascribed to accretion disk precession (Ogilvie \&
Dubus 2001).  Kaaret \& Feng (2007) argued that the 62 day period in
X41.4+60 was unlikely to be a superorbital modulation because the
observed period is only consistent with the observed superorbital
periods of neutron star systems, but that the extreme luminosity of
X41.4+60 comfortably rules out an accreting neutron star (see their
Figure 4 for a distribution of observed superorbital periods). Similar
arguments would appear valid for the 115 day modulation in NGC 5408
X-1.  The 115 day period is shorter than all known superorbital
periods for black hole candidate binaries, and its peak luminosity is
well in excess of the Eddington limit for a neutron star ($\approx
10^{38}$ ergs s$^{-1}$).

Using X-ray timing measurements with {\it XMM-Newton}, Strohmayer et
al. (2009) have recently shown that the strong QPO observed in NGC
5408 X-1 varies in frequency and amplitude with changes in the X-ray
flux and energy spectrum in a manner that closely mimics the
correlated temporal and spectral variations observed in stellar mass
black holes in the so-called Intermediate State (also known as the
Steep Power-law State).  Moreover, recent optical spectroscopy
reported by Kaaret \& Corbel (2009) indicates that its optical
counterpart has a significant contribution from reprocessing of the
X-ray luminosity in the outer parts of the disk.  The so-called ``slim
disks'' that may exist at super-Eddington accretion rates
preferentially radiate more flux out along the disk axis, and
therefore have relatively less available for reprocessing (Abramowicz
et al 1988). These models thus have difficulty accounting for the
observed optical to X-ray flux ratio in NGC 5408 X-1.  Further,
modeling of the observed nebular optical emission lines supports the
conclusion that NGC 5408 X-1 radiates $\sim 10^{40}$ ergs s$^{-1}$ in
an approximately isotropic manner (Kaaret \& Corbel 2009). These
observations support the notion that NGC 5408 X-1 harbors a
``standard'' thin, optically thick accretion disk similar to those
inferred to exist in Galactic black hole binaries.

The 115 day modulation we have found in NGC 5408 X-1 shows interesting
energy dependent effects that would appear consistent with orbital
modulation.  A system which shows qualitatively similar effects to
those seen in NGC 5408 X-1 is Cyg X-1. Specifically, the orbital
modulation in Cyg X-1 has been modeled in the context of absorption
and scattering of X-rays in the partially ionized wind from the
companion star (Wen et al. 1999). This model predicts a decrease in
modulation amplitude with increasing energy, and a minimum in the
hardness ratio at the maximum of the modulation profile (see Wen et
al. 1999, Figure 3), both of which are evident in the NGC 5408 X-1
data. While the shapes of the modulation profiles in Cyg X-1 do not
exactly match the results we find for NGC 5408 X-1, the behavior of
the $E > 1$ profile (see Figure 3) appears qualitatively similar.  An
important distinction would seem to be that in the case of NGC 5408
X-1 we can directly observe the thermal disk flux whereas the orbital
modulation measurements for Cyg X-1 concern the low-hard state, when
presumably its thermal flux is weak or absent. Moreover, any disk flux
appears largely shortward of the RXTE/PCA energy band.  Cyg X-1 is
also a wind accreting system, whereas wind accretion would likely be
unable to account for the high luminosity of NGC 5408 X-1.  

If the 115 day period is indeed the orbital period of a Roche lobe
filling binary, then the mean density of the secondary is constrained
to be $\rho_{mean} \approx 0.2 (P_{\rm days})^{-2}$ g cm$^{-3}$ (Frank
et al. 2002). For a period of 115.5 days this gives $\rho_{mean} =
1.50 \times 10^{-5}$ g cm$^{-3}$, and the companion would have to be a
giant or supergiant star.  Recent observations suggest that a
substantial fraction of the optical counterpart is due to reprocessing
of the X-ray flux in an accretion disk, and while the companion star
still has not been observed directly, the recent optical measurements
suggest that it is probably a giant in the 3 - 5 $M_{\odot}$ range,
and with a spectral type of B or later (Kaaret \& Corbel 2009).
Binary evolution calculations for IMBHs and massive companions appear
consistent with the notion that NGC 5408 X-1 contains a $\sim 1000
M_{\odot}$ black hole with a 3 - 5 $M_{\odot}$ companion (Li 2004).
Continued monitoring of NGC 5408 X-1 with {\it Swift} is important in
order to confirm the orbital nature of the modulation.

\clearpage

\begin{figure}
\begin{center}
 \includegraphics[width=6.5in, height=5.5in]{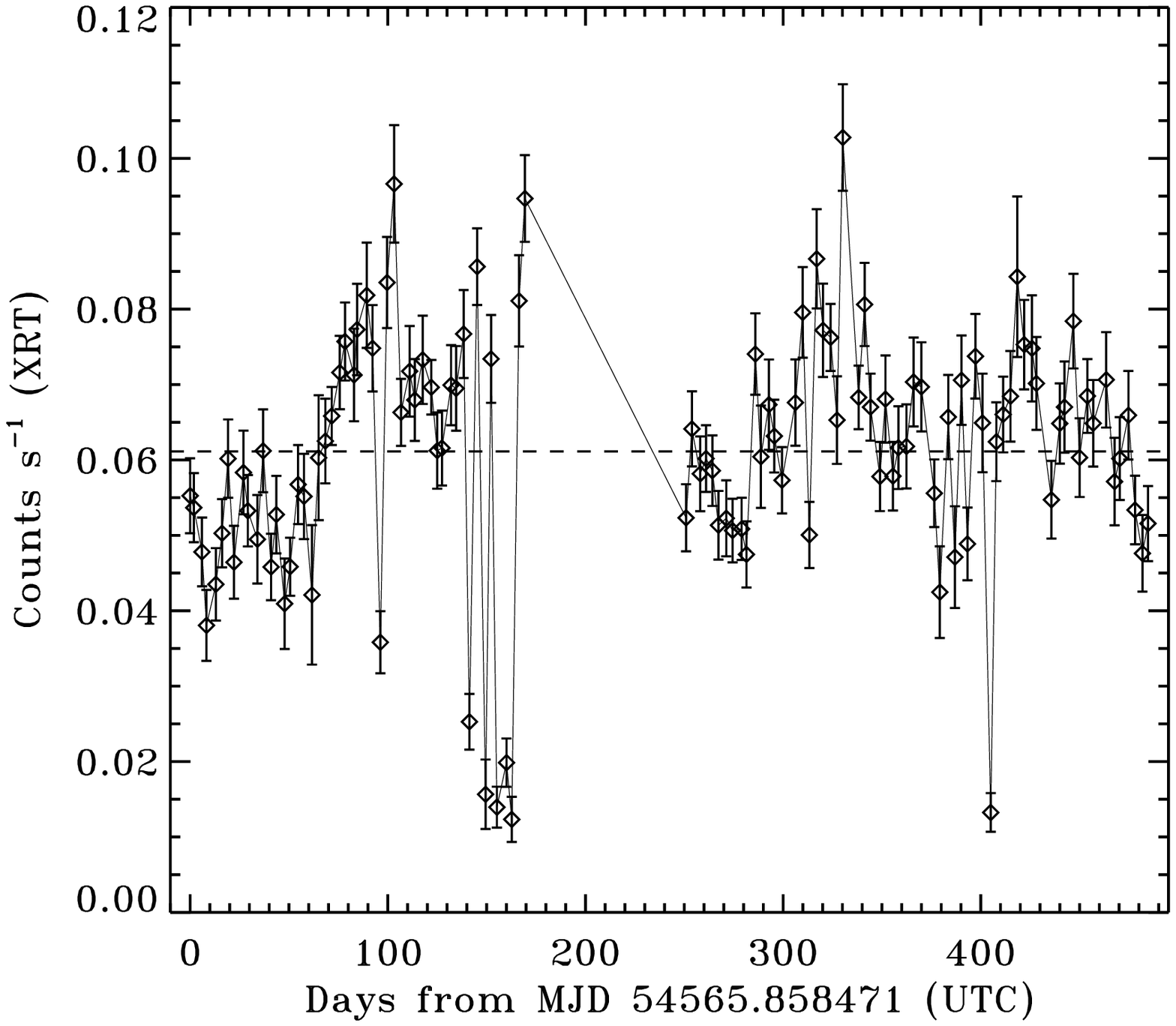}
\end{center}
Figure 1: X-ray lightcurve of NGC 5408 X-1 in the 0.2 - 8 keV band
derived from {\it Swift}/XRT measurements. A count rate of 0.06
s$^{-1}$ corresponds to a 0.3 - 10 keV unabsorbed flux of $4 \times
10^{-12}$ ergs cm$^{-2}$ s$^{-1}$. The mean count rate of 0.061
s$^{-1}$ is marked by the dashed horizontal line. Time zero
corresponds to MJD 54565.85847 (UTC).
\end{figure}
\clearpage

\begin{figure}
\begin{center}
 \includegraphics[width=6.5in, height=5.5in]{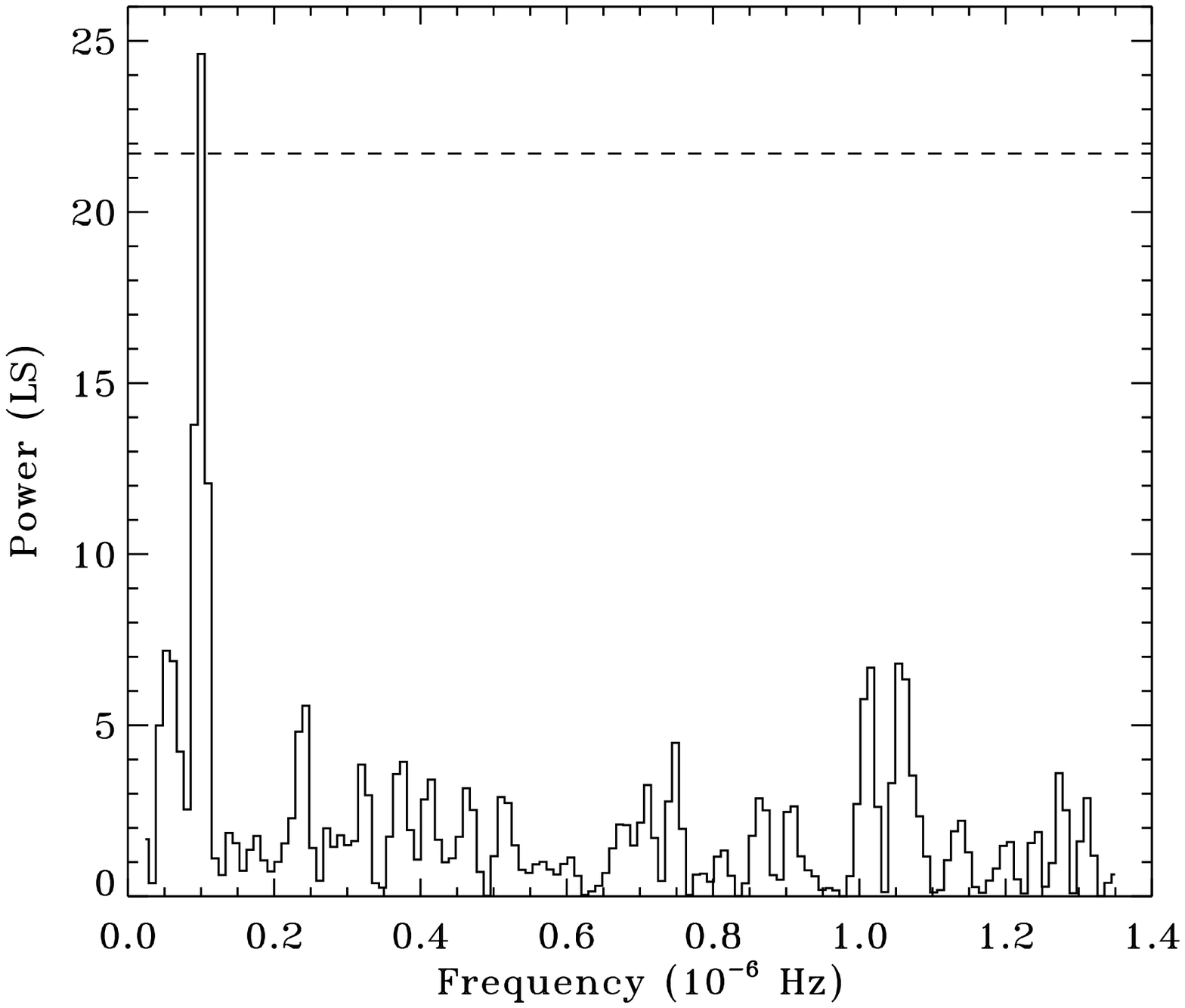}
\end{center}
Figure 2: Lomb-Scargle periodogram computed from the {\it Swfit}/XRT
light curve of NGC 5408 X-1.  There are 140 independent frequency bins
plotted with a resolution of 9.54 {\it nano}Hz.  The Nyquist frequency
is 1.35 $\mu$Hz (corresponding to a period of 8.6 days).  The
horizontal dashed line denotes the $3\sigma$ detection level. The peak
value of 24.7 occurs at a frequency of $1.0018 \times 10^{-7}$ Hz (a
period of 115.5 days).
\end{figure}

\clearpage

\begin{figure}
\begin{center}
 \includegraphics[width=6.5in,height=5.5in]{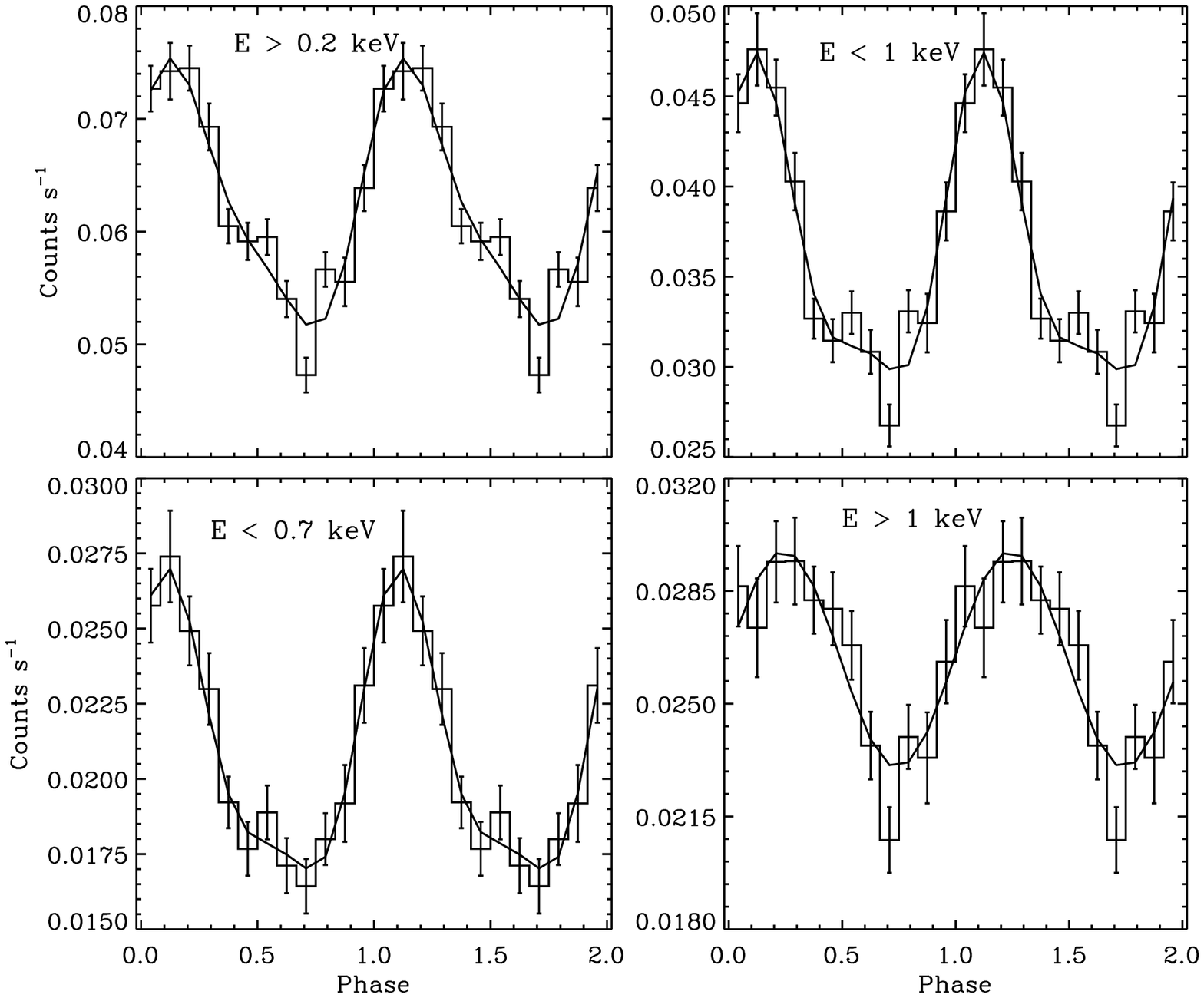}
\end{center}
Figure 3: X-ray flux from NGC 5408 X-1 in different energy bands
folded at a period of 115.5 days. Profiles were computed with 12 phase
bins, and two cycles are shown for clarity. Proceeding clock-wise from
the upper left panel the energy bands are, $0.2 < E < 8$ keV; $0.2 < E
< 1$ keV; $1 < E < 8$ keV; and $0.2 < E < 0.7$ keV. The best fitting
model with two Fourier components is also plotted in each panel.  Note
the change in profile amplitude and shape with energy. See Table 1 for
a summary of the model fits, and the text for additional discussion.
\end{figure}

\clearpage

\begin{figure}
\begin{center}
 \includegraphics[width=6.5in,height=5.5in]{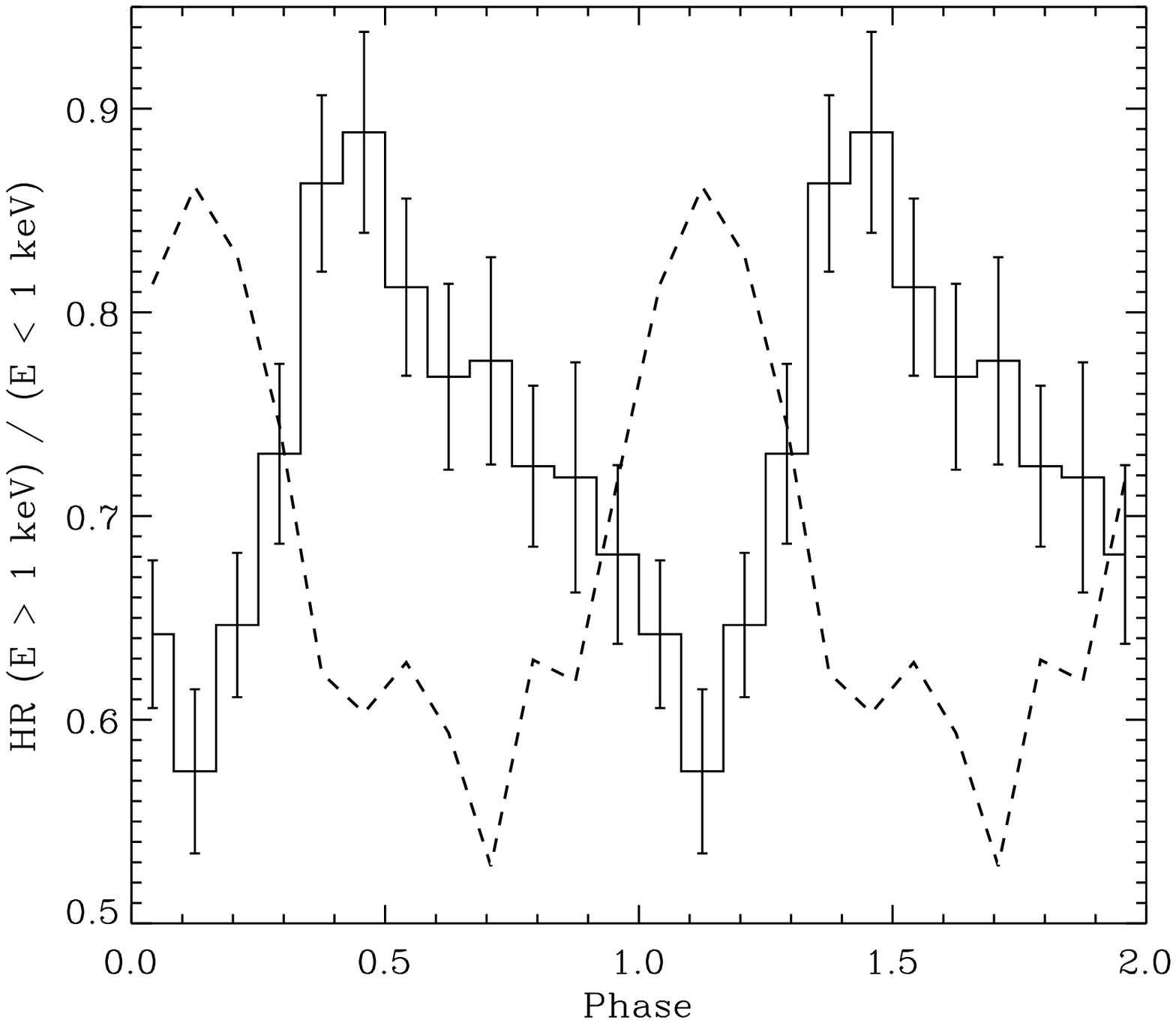}
\end{center}
Figure 4: Hardness ratio as a function of phase of the 115.5 day
modulation in NGC 5408 X-1.  The hardness ratio is defined as the
ratio of count rates for photons with energy $> 1$ keV to that for
photons with energy $< 1$ keV.  We also plot the full-band (0.2 - 8
keV) folded profile for comparison (dashed curve).
\end{figure}

\clearpage

\begin{deluxetable}{ccccccc}
\tablecaption{Results of Pulse Profile Modeling for NGC 5408
X-1\tablenotemark{1}} 
\tablehead{\colhead{Profile} & \colhead{A ($10^{-2}$)} & 
\colhead{B ($10^{-2}$)} 
& \colhead{$\phi_0$} & \colhead{C ($10^{-2}$)} & \colhead{$\phi_1$} & 
\colhead{$f_{amp}$\tablenotemark{a}} } 

\startdata
 
$E > 0.2$ keV & $6.23 \pm 0.05$ & $1.10 \pm 0.08$ & 
$0.915 \pm 0.011$ & $0.27 \pm 0.07$ & $0.46 \pm 0.02$ & 
$0.182 \pm 0.013$ \\

$E < 0.7$ keV & $2.09 \pm 0.03$ & $0.48 \pm 0.05$ & 
$0.874 \pm 0.014$ & $0.14 \pm 0.04$ & $0.473 \pm 0.024$ & 
$0.238 \pm 0.022$ \\

$E < 1$ keV & $3.64 \pm 0.04$ & $0.83 \pm 0.06$ & 
$0.882 \pm 0.011$ & $0.27 \pm 0.06$ & $0.485 \pm 0.016$ & 
$0.241 \pm 0.017$ \\

$E > 1$ keV & $2.64 \pm 0.03$ & $0.34 \pm 0.05$ & 
$0.992 \pm 0.023$ & $0.096 \pm 0.05$ & $0.853 \pm 0.040$ &
$0.133 \pm 0.018$ \\

\enddata 

\tablenotetext{1}{Summary of fits to energy dependent pulse profiles 
for NGC 5408 X-1.  The fitted model is $I = A + B\sin[2\pi
(\phi - \phi_0)] + C\sin [4\pi (\phi-\phi_1)]$.   }

\tablenotetext{a}{Fractional modulation amplitude defined as $(B^2 +
C^2)^{1/2}/A$.}

\end{deluxetable}

\end{document}